\documentclass[11pt]{article}

\usepackage[utf8]{inputenc}
\usepackage[T1]{fontenc}

\usepackage{amsmath,amssymb,amsthm,bbm}

\usepackage{graphicx,transparent,eso-pic}

\usepackage[document]{ragged2e}
\usepackage{pagecolor,color}
\usepackage{soul}

\usepackage{ifthen}

\usepackage{natbib}

\usepackage{url}

\usepackage[margin=2cm]{geometry}

\usepackage{threeparttable}
\usepackage{mdframed}
\usepackage{listings}

\title{Space Matters: extending sensitivity analysis to initial spatial conditions in geosimulation models}

\author{%%%% Author details
J. Raimbault$^{1,2,\ast}$, C. Cottineau$^{3,4}$, M. Le Texier$^{5}$, F. Le N{\' e}chet$^{6}$, R. Reuillon$^{1}$\medskip\\
$^{1}$UPS CNRS 3611 ISC-PIF, Paris, France\\
$^{2}$UMR CNRS 8504 G{\'e}ographie-cit{\'e}s, Paris, France\\
$^{3}$Centre for Advanced Spatial Analysis, University College London, UK\\
$^{4}$UMR CNRS 8097 Centre Maurice Halbwachs, Paris, France\\
$^{5}$UMR 6266 IDEES, Universit{\'e} de Rouen Normandie, France\\
$^{6}$Universit{\'e} Paris-Est, Laboratoire Ville Mobilit{\'e} Transport, Marne-la-Vallée, France\medskip\\
$^{\ast}$ Corresponding author: \texttt{juste.raimbault@polytechnique.edu}
}

\date{}

\begin{document}

\maketitle

\begin{abstract}
Although simulation models of geographical systems in general and agent-based models in particular represent a fantastic opportunity to explore socio-spatial behaviours and to test a variety of scenarios for public policy, the validity of generative models is uncertain unless their results are proven robust and representative of 'real-world' conditions. Sensitivity analysis usually includes the analysis of the effect of stochasticity on the variability of results, as well as the effects of small parameter changes. However, initial spatial conditions are usually not modified systematically in geographical models, thus leaving unexplored the effect of initial spatial arrangements on the interactions of agents with one another as well as with their environment. In this paper, we present a method to assess the effect of some initial spatial conditions on simulation models, using a systematic spatial configuration generator in order to create density grids with which spatial simulation models are initialised. We show, with the example of two classical agent-based models (Schelling's models of segregation and Sugarscape's model of unequal societies) and a straightforward open-source work-flow using high performance computing, that the effect of initial spatial arrangements is significant on the two models. Furthermore, this effect is sometimes larger than the effect of parameters' value change.

\medskip

\end{abstract}

%%%%%%%
%% Highlights
%
% - We investigate the sensitivity of geosimulation models to initial spatial conditions.
% - We generate stylized population density grids.
% - We assess the sensitivity of two classical agent-based models, Schelling's model and Sugarscape's model.
% - Strong quantitative and qualitative effects of spatial initial conditions are shown for these models.

%%%%%%%%%%%%%%%%%%%%%%
\section{Introduction}
%%%%%%%%%%%%%%%%%%%%%%

Computer simulation has been recognised and is increasingly used by geographers as an efficient tool to explore geographical processes, hypotheses and predictive scenarios within virtual laboratories \citep{batty1971modelling, batty2007model, carley1999generating, Quesneletal2009}. It has been identified as an emerging field and coined under the term geosimulation by \cite{benenson2004geosimulation}. Simulation also appears as a way to overcome the difficult analytic resolution of many spatial models which were developed in the past, as well as to explore the possible (alternative) trajectories of path-dependent social and ecological systems. The specificity of geographical models compared to other social science models is that space and spatial interactions are given a prime role, geographers being driven by an explicit interest in studying the way space influences the outcomes of social processes modelled. We think that simulation approaches are uniquely positioned to represent the complexity of socio-spatial interactions, provided that models include relevant spatial descriptions and behavioural rules which take space into account, and provided that the model evaluation includes a sensitivity analysis of the model outputs to the way space is represented. Unfortunately, the first condition is not always met, and the second is seldom even mentioned. This paper aims to fill a methodological and conceptual gap, which is a systematic testing of the sensitivity of a model's outcomes to its initial spatial conditions. To demonstrate the genericity of our approach, we develop two applications with classic simulation models commonly used as case studies for comparing and aligning simulation models \citep{Axtelletal1996, wilensky2007making}: \citet{schelling1971dynamic}'s model of segregation and \citet{EpsteinAxtell1996}'s Sugarscape model.\\

\subsection{Definition of the problem}

Geographical systems can be crudely described as social agents interacting with one another, within a limited portion of space. Social agents thus constitute the microscopic level of the system, and they are framed by a system-time that evolves irreversibly, creating temporal and cumulative effects, also known as path-dependency \citep{arthur1994increasing}. Therefore, observing a system at different points in time does not equate to observing different systems at a single point in time. This general property of ergodicity applies to geographical elements such as road networks or built-up areas \citep{pumain2003approche}. Similarly to what \citet{gell1995quark} calls \emph{frozen accidents} in complex systems generally, a given configuration contains clues about past bifurcations, that can have had dramatic effects on the state of the system. Therefore, strong spatio-temporal path-dependencies in the trajectory of individual cities and changing social environments over time prohibits the use of ergodic models. Ironically, these very models tend to be the models most frequently used in geosimulation.

Self-organization has been shown to be a central feature of geographical systems in general and of cities in particular ~\citep{AllenSanglier1981,saint1989villes, Portugali2000}. In the vocabulary of complex systems, cities also exhibit emergent properties at macroscopic scales~\citep{pumain2006hierarchy, AzizAlaouiBertelle2009}, which can be simulated through microscopic interactions between agents~\citep{Wu2002, Batty2007}. Complexity is partially due to bifurcations, which are determinant in spatial systems~\citep{Wilson1981, Wilson2002}. Indeed, in spatially explicit simulation models, the non-linearity of local interactions is very likely to sublimate small perturbations in the initial spatial setting, making it difficult to interpret the resulting global structures. In that sense, the impact of initial \emph{spatial} settings on final outcomes is assumed to be significant just as any other initial conditions, but of more interest to the geographer. \\

Finally, although this may seem obvious, cities are not regular grids, and the distribution of density (of jobs, residents, buildings, etc.) is far from isotropic, even in sprawled cities. On the contrary, there is a significant diversity in the way people, activities and structures are distributed in cities. In Europe for example, \citet{LeNechet2015} quantifies and classifies six broad types of residential density distributions. However, most geographical models, especially cellular automata, still represent cities as uniform grids of isotropic density. Even in applied cases when GIS geometries of a particular city are used, the spatial distribution of agents tend to be approximated by a constant density \citep{arribas2014diverse}, although previous research shows that it is computationally and methodologically feasible to use accurate locations in a simple model such as Schelling's \citep{benenson2002entity}. The isotropic simplification is potentially harmful to the representation of urban processes because density and accessibility have environmental, economic and social consequences. Additionally, we expect the initial spatial distribution of agents to influence simulation results in the long run \citep{Castellanoetal2009}, because the agents' rule of action itself may depend on the spatial structure of the environment. For example, households can have different preferences with respect to the built-environment they might want to live in \citep{SpielmanHarrison2014}, or agents moving around will sense a different set of objects within the same fixed radius depending on the topology \citep{Banos2012} and distribution of density of the sensed environment \citep{LauriJaggi2003, FossettDietrich2009}. The way modellers represent initial space is therefore a central element of geographical simulation models. However, this step is rarely explicit. A meaningful way of representing initial space might be to consider, not necessarily the peculiarities of every city, but at least their broad density structures so as to estimate the variability of the model behaviour to different plausible spatial arrangements.

\subsection{Objective}

In this paper, we offer a methodological solution to the problem of sensitivity of simulation models to initial spatial conditions with two application cases (Schelling and Sugarscape). In no way do we pretend to provide a full exploration of these two particular models, their attractors and/or potential policy implications. Instead, we present a way of performing a sensitivity analysis to initial spatial conditions of models by using a spatial generator to produce a variety of density grids, which are taken as input by simulation models at initialisation. The generator being controlled by its own parameters, we can then relate the parameters used to generate initial spatial conditions to the variation simulation outcomes. The purpose is two-fold: (i) to test the robustness of simulation results to small variations of generator parameters and (ii) to study the non-trivial effects of typical categories of spatial distribution (monocentric \textit{vs.} polycentric for example) on the results of a given model. Our approach allows for a systematic comparison of several aspects of the spatial configuration problem, which have been suggested by \citet{filatova2013spatial}, but hardly implemented and achieved in previous studies to our knowledge. In particular, it is applied to the effects of urban form on simulation results, using Schelling's model as a first case study and Sugarscape as a second one. 

%%%%%%%%%%%%%%%%%%%%%%
\subsection{Previous considerations on the effects of the spatial configuration in simulation models}

In 'real-life', different spatial configurations of people in a city tend to be associated with different distributions of income, carbon emissions, educational outcomes, etc. For example, \citet{wheeler2006urban} shows that, in the US, sprawling cities are more unequal than their compact counterparts with respect to income. Dynamically, sprawl in American cities consists in the addition of new developments which have been occupied by different groups of population, resulting in a concentration of the wealthy in suburban pockets and of pockets of poverty in the inner city area \citep{jargowsky2002sprawl}. Similarly, in terms of pollution for example, \citet[p.173]{schwanen2001travel} show that \textit{``deconcentration of urban land uses encourages driving and discourages the use of public transport as well as cycling and walking''}. The discussion of these effects has also been associated with the field of geostatistics since the exposure of the Modifiable Areal Unit Problem (MAUP) \citep{Openshaw1984, FotheringhamWong1991}. For example, \citet{Kwan2012} has argued for a careful examination of what she coins the 'uncertain geographic context problem' (UGCoP), i.e. of the spatial configuration of geographical units even when the size and delineation of the area are the same.\\

Considerations of such issues in the geosimulation literature are rather scarce. However, there have been some noticeable attempts at analysing the impact of three types of initial spatial characteristics on model outcomes:
\begin{itemize}
\item The accuracy of geo-localised input data;
\item The shape, precision and boundaries of the modelled spatial system;
\item The degree of spatial heterogeneity modelled.
\end{itemize}

\subsubsection{Geo-localised input data accuracy}

\citet{Thomasetal2017} show that data selection in LUTI model is inter-related with the delineation of the spatial system boundaries and the scale of analysis. They provide a few examples on how the use of Exploratory Spatial Data Analysis (ESDA) prior to simulation runs can help avoiding measurement errors of model behaviour and outcomes. In the context of spatial interaction models, \citet{hagen2012new} acknowledge the dilemma between spatial resolution and the computational burden, and suggest a method of adaptive zoning (where the size of destination zones depends on the distance to origin) to solve it.

\subsubsection{Spatial system shape, precision, and boundaries} 

\citet{Axtelletal1996} highlight the sensitivity of the average number of stable cultural regions generated to the effect of the territory width implemented in a version of the Sugarscape model which is docked (i.e. made equivalent to) to Axelrod Culture Model. \citet{FlacheHegselmann2001} show that chances for random emergence of a stable cluster of similar agents in a Schelling-like model are higher in a rectangular grid and lower in a hexagonal grid and that an irregular (Voronoi-diagram) city lattice structure favours migration stabilisation around decentralised clusters of similar agents. \citet{Banos2012} compares the behaviour of Schelling segregation model on city lattices formalized as either grid, random, scale-free and Sierpinski networks and concludes that the presence of cliques in graph-based urban structures favours segregationist behaviours. \citet{LeTexierCaruso2017}, using a set of different theoretical spatial systems, demonstrate the impact of the regularity and aggregation levels, or centrality/periphery effects, on spatial diffusion dynamics of euro coins. Similar issues were also dealt with in physical sciences: for example, \cite{horritt2001effects} studies the effects of grid cell size on the behaviour of a raster flood model and shows that increasing resolution does not increase model prediction performance below a certain level. Similar conclusions are obtained by \cite{vazquez2002effect}, unveiling an intermediate optimal spatial resolution regarding model performance and computation time. Spatial resolution also plays a role in the Schelling model: \citet{Singhetal2009} show that the segregation patterns for certain tolerance values are strictly a small city phenomenon (8x8 city-lattice) and do not work for a larger spatial lattice (100x100), where segregation appears only for certain combinations of tolerance threshold and vacancy density values.

\subsubsection{Spatial heterogeneity}

\citet{StaufferSolomon2007} introduce asymetric interactions and empty residences in Schelling's model run on a large and regular lattice. They reveal conjoint and non-linear effects on the vacancy rates and tolerance levels on segregation patterns. \citet{Gauvinetal2010} run Schelling's segregation process in an open city-lattice to study how the variations in tolerance levels, vacancy rates and city attractiveness may create lines of vacancy lots between clusters of agents. They conclude on the functional role of vacancies, which allow weakly tolerant agents to live and be satisfied in a city environment they nevertheless perceive as hostile. \citet{HatnaBenenson2012} show that their model replications run on a 50x50 torus with 2\% of empty cells were not sensitive to the initial patterns (random and fully segregated distribution of agents). In ecology, \citet{smith2002predicting} study the spread of a disturbance in an heterogeneous landscape using a percolation model, and show that landscape structure has a significant influence on final patterns of contamination outcome. In a spatial epidemics model for an infectious disease, parameterised on real data, \citet{smith2002predicting} finds that the physical landscape heterogeneity, in particular the presence of rivers, locally influence the propagation speed.

Generally, sensitivity analyses have focused either on the spatial context (extent of the environment, shape of the zones if applicable - squares, hexagons, etc., objects - links, rasters, etc.) or on the spatial encoding of the heterogeneity of space (algorithms of disaggregation, interpolation, etc.), but rarely on both at the same time. More so, the references cited in this section are few compared to the mass of geosimulation articles published without any mention of dependence of output on initial spatial condition.

%\cite{ligmann2013spatially}

In this paper, we build on these contributions and go further by systematically measuring the impact of the city density distribution aggregate model behaviours (e.g. the final segregation level for Schelling). We illustrate the potential generalisation of our approach by running two distinctive agent-based models: Schelling's model and Sugarscape.

%%%%%%%%%%%%%%%%%%%%%%
\section{Methods}
%%%%%%%%%%%%%%%%%%%%%%

\begin{figure}[!t]
\centering
\includegraphics[width=0.9\textwidth]{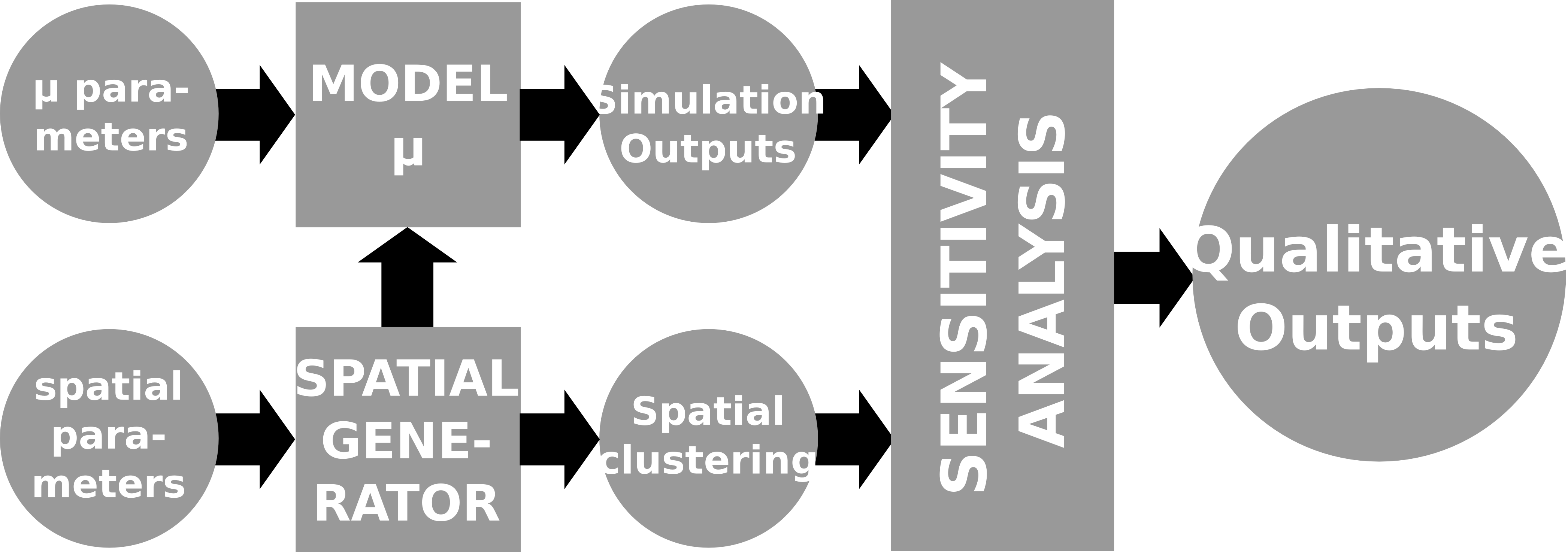}
\caption{\textbf{General workflow.}} \label{fig:method}
\end{figure}

The general method workflow of our method is illustrated in Figure \ref{fig:method}. In addition to the usual protocol (upper branch of the figure), which consists of running a model $\mu$ with different values of its parameters, we introduce a spatial generator which depends on its very own set of parameters and feeds the model $\mu$ at initialisation (lower branch). We call these parameters ``generator parameters'' to distinguish them from the standard parameters of the models and to highlight their higher position in the simulation process. The resulting configurations \emph{can be} clustered into qualitative types of spatial patterns. The sensitivity analysis relates the variations in the model's outcomes to how the density spatial distribution was generated and to the patterns of density generated. In particular, we want to emphasize that spatial effects derive not only from grid size or shape effects, but also from the heterogeneity of the distribution of geographical entities (people, housing, networks, etc.) in space. In the models we used as examples, the initial spatial configurations can be either flat or heterogeneous, monocentric or polycentric, based on external databases and on internal modelling (generation of synthetic population for instance \citep{bhat1999activity}).
%%%%%%%%%%%%%%%%%%%%%%
%%%%%%%%%%%%%%%%%%%%%%

In order to test the influence of initial spatial conditions on model outputs, we use a systematic method to compare \emph{phase diagrams}. Following \citet{Gauvinetal2009}, we define a phase diagram as the vector of final aggregated model outputs considered as a function of model parameters. We have as many phase diagrams than we have spatial grids, which makes a qualitative visual comparison not realistic (with around 50 different spatial configurations for each model experiment). A solution is to use systematic quantitative procedures to compare them to a reference case. Technically, because of stochasticity, we represent the output of the model for a given combination of parameter values $\vec{P}$ as the mean of the final values of an output indicator $O$ obtained for the replications of the model initialized with $\vec{P}$.\\

To our knowledge there exists no single well established method to compare phase diagrams in the agent-based modelling and geosimulation literature. We introduce a measure of the relative distance $d_r$ between two phase diagrams $\mu_{\vec{\alpha}_1}$ and $\mu_{\vec{\alpha}_2}$. Phase diagrams are denoted by the same function $\mu$ indexed by the generator parameters $\vec{\alpha}$, which capture the spatial configuration (in practice these can be parameters of an upstream model to generate the configuration, or a description of the configuration itself). The $d_r$ measure is taken as the share of between-diagrams variability relative to their internal variability, given formally in the case of a one-dimensional phase diagram by

\begin{equation}\label{eq:phase-distance}
d_r\left(\mu_{\vec{\alpha}_1},\mu_{\vec{\alpha}_2}\right) = 2 \cdot \frac{d(\mu_{\vec{\alpha}_1},\mu_{\vec{\alpha}_2})^2}{Var\left[\mu_{\vec{\alpha}_1}\right] + Var\left[\mu_{\vec{\alpha}_2}\right]}
\end{equation}

where $\mu_{\vec{\alpha}_i}$ are the phase diagrams at given generator parameters $\vec{\alpha}_i$. $d$ is a functional distance that we take simply as Euclidean distance. The internal variabilities are estimated as the variance within each phase diagram $\mu_{\vec{\alpha_i}}$. For a multi-dimensional phase diagram, we average these relative distances over the components. Given a set of phase diagram to be compared, we will study the distribution of this distance to an arbitrary phase diagram for all diagrams, rather than an aggregated measure which would be similar to global sensitivity methods \citep{saltelli2008global}.

The last methodological point which we need to emphasize is the relationship between the present workflow and model exploration workflows in general. The ideas of multi-modelling and extensive model exploration are nothing from new - ~\cite{openshaw1983data} already advocated for ``model-crunching'' in 1983 -, but their effective use only begins to emerge thanks to the development of new methods and tools together with an explosion of computation capabilities. The model exploration platform OpenMOLE~\citep{reuillon2013openmole} allows to embed any model as a blackbox, to write flexible exploration workflow using advanced methodologies such as genetic algorithms and to distribute transparently the computation on large scale computation infrastructures such as clusters or computation grids. In our case, this tool is a powerful way to embed both the sensitivity analysis and the sensitivity analysis to initial spatial conditions, and to allow the coupling of any spatial generator with any model in a straightforward way as long as the model can take its spatial configuration as an input or from an input file. In this paper, we use the OpenMOLE platform for the spatial environment and the model coupling, placing ourselves in the framework of multi-modelling \citep{cottineau2015modular}. We use therefore OpenMOLE's functionalities for model embedding through workflow, design of experiments (parameter sampling) and high performance environment access.

%%%%%%%%%%%%%%%%%%%%%%
\subsection{Spatial generator of density grids}

\begin{figure}[!t]
\centering
	\includegraphics[width=\textwidth]{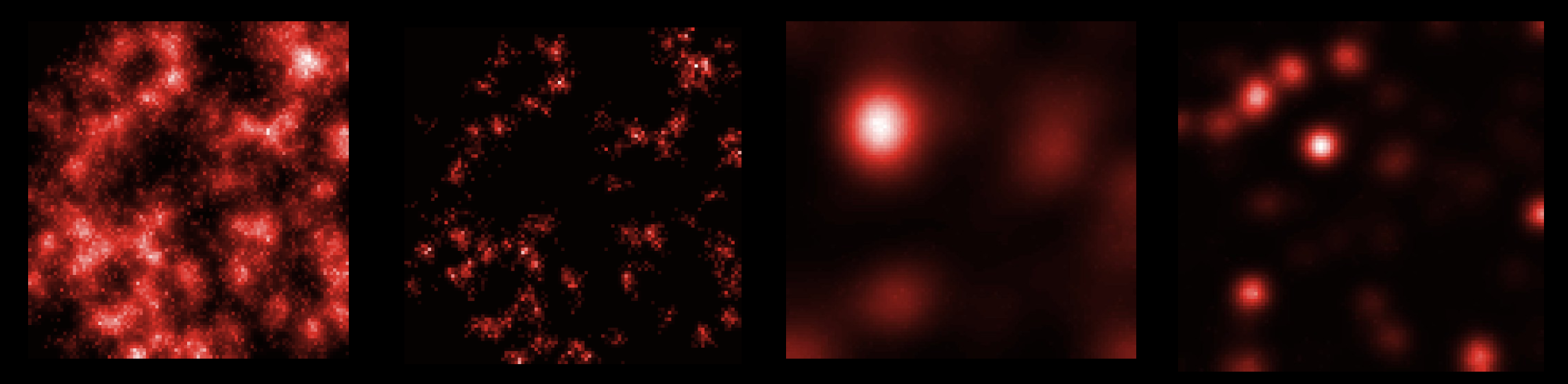}
 \caption{\textbf{Four examples of grids produced by the spatial generator.} The lighter the red, the denser the area. Changing the growth rate $N$ allows to have more or less chaotic shapes (two first compared to the two last grids for example) corresponding to different levels of convergence of the model, whereas local radius can be tuned with the interplay of aggregation strength $\alpha$ and diffusion strength $\beta$. Parameter values used here are for the first grid $(\alpha = 0.4,\beta=0.006, N = 25, n_d = 1, t_f = 971)$, the second $(\alpha = 0.4,\beta=0.006, N = 25, n_d = 1, t_f = 176)$, the third $(\alpha = 1.4,\beta=0.045, N = 102, n_d = 2, t_f = 618)$, and the fourth $(\alpha = 1.8,\beta=0.114, N = 108, n_d = 1, t_f = 227)$.}
\label{fig:spatialGen}
\end{figure}

The spatial generator applies an urban morphogenesis model \citep{Batty2007} which has been generalised, explored and calibrated by~\citet{raimbault2018calibration}. An open implementation and a characterisation of the urban forms which the model can produce allow us to integrate it easily into our workflow. Grids are generated through an iterative process which, starting from an empty grid, adds a quantity $N$ (population) at each time step $t$, allocating it through preferential attachment on population density, characterised by its strength of attraction $\alpha$. More precisely, each added unit has a probability equal to $P_i^{\alpha}/\sum_k P_k^{\alpha}$ to be added to a patch $i$ with population $P_i$, all $N$ units being added independently and in parallel. At the end of each time step, this growth process is smoothed $n_d$ times using a diffusion process of strength $\beta$: each patch transmits an equal share of $\beta\cdot P_i$ to its Moore neighborhood (i.e. its 8 surrounding patches). To avoid border effects such as a reflexion on the border of the world, border patches diffuse to the outside. The procedure stops when a fixed number of steps $t_f$ is reached. The grid then has a population of $t_f \cdot N$ (the population lost due to diffusion process to the outside is reallocated through a normalization procedure at the end of the steps). Grids are thus generated from the combination of the values of these four generator parameters $\alpha$, $\beta$, $n_d$ and $N$, in addition to the random seed. To ease our exploration, only the distribution of density is allowed to vary rather than the size of the grid, which we fix to a 50x50 square environment. We furthermore fix the total population at $t_f\cdot N = 100,000$, and determine therein the number of steps needed at a given $N$. Typical value ranges for the  parameters will be taken as, following \citet{raimbault2018calibration}, $\alpha\in\left[0.5,4.0\right]$, $\beta \in\left[0,0.3\right] $, $N\in \left[100,10000\right]$, $n_d\in\left[1,4\right]$. We illustrate in Fig.\ref{fig:spatialGen} the variety of spatial configurations that can be generated.

%%%%%%%%%%%%%%%%%%%%%%
%%%%%%%%%%%%%%%%%%%%%%

% Generator parameters values:
%  1) ex_sp-diffusion=0.0060_sp-growth-rate=25_sp-diffusion-steps=1_sp-alpha-localization=0.4_ticks=971_sp-population=24275.000000000102
%  2) ex_sp-diffusion=0.0060_sp-growth-rate=25_sp-diffusion-steps=1_sp-alpha-localization=0.4_ticks=176_sp-population=4400.000000000003
%  3) ex_sp-max-pop=80260_sp-diffusion=0.045_sp-growth-rate=102_sp-diffusion-steps=2_sp-alpha-localization=1.4_ticks=618
%  4) ex_sp-max-pop=382170_sp-diffusion=0.114_sp-growth-rate=108_sp-diffusion-steps=1_sp-alpha-localization=1.8_ticks=227

\begin{figure}[!t]
\centering
	\includegraphics[width=0.9\textwidth]{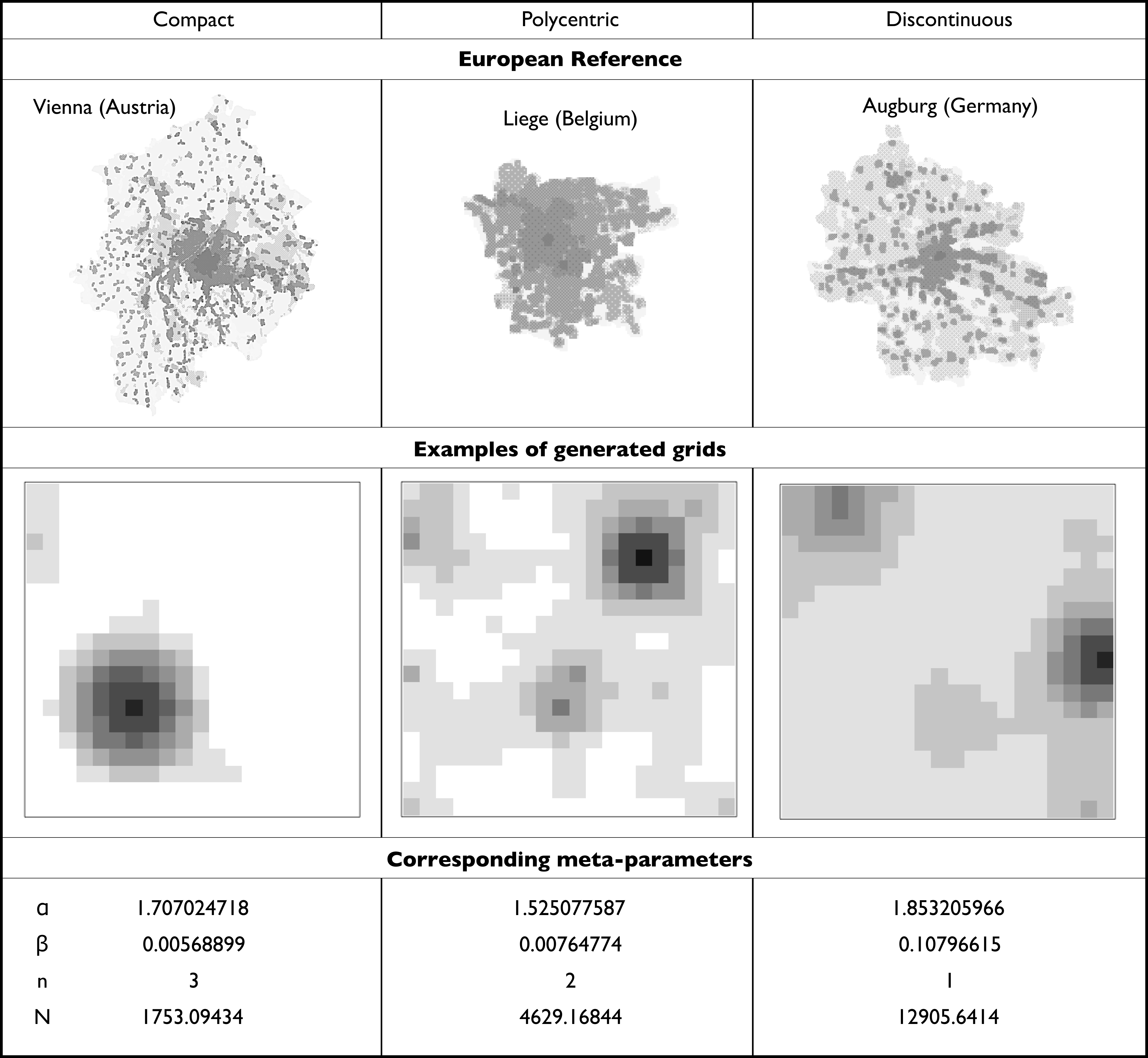}
 \caption{Correspondence between European urban density structures and grids produced with the spatial generator.} 
\label{fig:densityTypes}
\end{figure}

In order to generate density grids which correspond to empirical density distributions, we select among the generated grids using an objective function which matches the point cloud of 110 metropolitan areas in Europe described by four dimensions of spatial structure: their concentration index, hierarchy index, centrality index and continuity index (cf. \cite{LeNechet2015}). We sample the generator parameter space using a Latin Hypercube Sampling, which is a convenient technique to have a scatter with high discrepancy. We sample 2000 points in the 4-dimensional space of parameters {$\alpha$, $\beta$, $n_d$, $N$}. It yields a subset of 170 grids matching empirical densities, which constituted our set of different initial spatial conditions. These are further clustered into three classes of morphology (figure \ref{fig:densityTypes}): compact (e.g. Vienna), polycentric (Liege) and discontinuous (Augsburg). This clustering allows to evaluate the non-trivial effects of a meaningful urban form on simulation results. We select 15 grids of each type to work with in the computation of sensitivity analysis of an intra-urban model (cf. section \ref{sec:qualResults}).

The spatial generator and its resulting grids are relevant to the case study models we have picked (Schelling and Sugarscape) because it produces density grids at a ``metropolitan scale'', which is the scale at which both model were initially intended to be. In the case of Schelling's segregation model for example, this scale is the one at which most empirical segregation indexes are computed and compared to the model outputs. In the case of Sugarscape, it corresponds to the whole city if the model is a metaphor for city resources \citep{batty2005agents}, or to a generic landscape where a resource is grown otherwise. In both cases, our point is that there exist many different patterns of density distribution in resource location and urban density and that acknowledging this diversity might leads to a variation in the model outputs. Furthermore, in urban models, we argue that the hypothesis of isotropic density is potentially the most unrealistic one, although unfortunately the most common one in Schelling implementations.

%%%%%%%%%%%%%%%%%%%%%%
%%%%%%%%%%%%%%%%%%%%%%

In the following section, we briefly recall the main components of the two ``classical'' agent-based simulation models used to test how spatial density variations may impact simulation models behaviour and results, and how general the method is.

\subsection{Case study models}

Schelling's model consists in an abstract urban housing market where agents of different attributes (for example: red or green) sense their environment, evaluate their satisfaction in terms of neighbourhood composition (how many reds and greens?), and relocate if unsatisfied. It has been shown by \cite{Schelling1969} that even tolerant agents tend to produce segregated patterns because of the complexity of their local interactions and the snowball effect of individual moves on the global distribution of agents in the city. The main parameters of this model are the tolerance level (maximum \% of agents different to {\it ego} accepted in the neighbourhood), the scope of sensing, the global majority/minority split and the percentage of vacant spaces in the housing market. In addition, we are interested in testing the impact of the initial spatial distribution of housing capacity in this project, using the generated grids. The outcome of the model is measured by a combination of three segregation indices: Dissimilarity, Moran's I and Entropy~\citep{brown2006spatial}. We use a ad-hoc implementation of the Schelling model, both in Scala for performance reasons and in NetLogo to ensure visualization of model dynamics. The pseudo-code of the implemented model is available in~\ref{app:code} and source code are available on the repository of the project.

In general, the implementations of Schelling models allow only one agent per cell, and their initial distribution is random, therefore following a uniform distribution across the modelled city. In this experiment, we allow more than one agent to be in a given cell. The potential density of a cell is defined by the density grid generated. If the potential density of a cell is not reached at initialisation, more agents can move into the cell during the course of the simulation, otherwise it is deemed full and unavailable for movers. The satisfaction and segregation indices are computed with regard to the people in the cell and the people present in neighbouring cells. Empirical distributions of density in cities are important in our framework because we want to test models with realistic ranges of initial patterns of density distribution. Therefore we cannot limit ourselves an isotropic square city in Schelling. We chose to use the actual distributions of European cities to constraint our density generation.

Sugarscape is a model of resource extraction which simulates the unequal distribution of wealth within a heterogenous population \cite{EpsteinAxtell1996}. Although it "is designed to study the interaction of many plausible social mechanisms" \cite[p.125]{Axtelletal1996}, we refer in this paper to the first (and simplest) version of the model, where "processes allow its agents to look for, move to, and eat a resource ("sugar") which grows on its $[...]$ array of cells". Agents of different vision scopes and different metabolisms harvest a self-regenerating resource available heterogeneously in the initial landscape, they settle and collect this resource, which leads some of them to survive and others to perish. The main parameters of this model are the number of agents, their minimal and maximal resource levels. In an urban environment, Sugarscape can be used to model how the spatial distribution of any type of goods or services can influence the spread of wealth among inhabitants. Following \cite{batty2005agents}, it can be considered as a metaphor of an urban system. We extend the implementation with agents wealth distribution of \cite{li2009netlogo}. The outcome of the model is measured by a Gini index of inequality for resource distribution. We are interested in testing the impact of the spatial distribution of the resource, using the generated grids.

\subsection{Experiment design}
For Sugarscape, we explore three dimensions of the parameter space: the total population of agents $P\in \left[10;510\right]$, the minimal initial agent resource $s_{-}\in \left[10;100\right]$ and the maximal initial agent resource $s_{+}\in \left[110;200\right]$. Each parameter is binned into 10 values, giving 1000 parameter points. We run 50 repetitions for each configuration, which yields reasonable convergence properties. The initial spatial configuration varies across 50 different grids, generated by sampling generator parameters in a LHS. We did not use the clustered grids to test the flexibility of our framework, which is demonstrated in this case by a direct sequential coupling of the generator and the model. Indeed, because the density distribution refers to the distribution of resource rather than to the representation of a city structure, we do not need the typology of urban density in this experiment. The full experiment thus equates to 2,500,000 simulations (1000 parameter combinations x 50 density grids x 50 replications). 

For Schelling's model, we also explore three dimensions of the parameter space of the model: the minimum proportion of similar agents required in the neighbourhood for the agent to be satisfied (or intolerance level) $S\in \left[0;1\right]$, the initial split of population, derived from the proportion of green population, $G\in \left[0;1\right]$ and the vacancy rate of the city $V\in \left[0;1\right]$. We sample 1000 parameter values using a Sobol sampling and run 100 repetitions for each configuration. We first try the same experiment design (50 density grids generated on the fly), then look at clustered grids representing urban densities. We choose 45 different grids among the ones which are most representative of the three types of urban morphology: 15 compact grids, 15 polycentric grids and 15 discontinuous grids. The last experiment thus equates to 4,500,000 simulations (1000 parameter combinations x 45 density grids x 100 replications). We use OpenMOLE to distribute the computation, and apply segregation measures to characterise the results.

As detailed in~\ref{app:convergence}, more repetitions are needed for Schelling indicators than for Sugarscape, in order to obtain a similar relative confidence in the estimation of averages. We ran for this reason a different number of replications for each model.

We choose different experiment designs, both for generator parameters and for the phase diagram, to demonstrate the robustness of the method to technical choices. In principle, our workflow applies regardless of the way we generate a spatial configuration (even taking real configurations) and the way we establish phase diagrams.

%%%%%%%%%%%%%%%%%%%%%%
\section{Results}
%%%%%%%%%%%%%%%%%%%%%%

The implementations of the models were done from NetLogo. We modified the sugarscape version with wealth of NetLogo model library (to be able to explore it intensively) and we implemented from scratch the Schelling model. Both pseudo-codes are available in~\ref{app:code}, and source code for models, grid classification and simulation results analysis is available on the open repository of the project at \url{https://github.com/AnonymousAuthor2/SpaceMatters}. Density grids are also available at this address. Simulation data are available for reproducibility on the dataverse repository at \url{https://doi.org/10.7910/DVN/A0N5GV}.

\subsection{Quantitative variation across grids}

We measure the distance of the phase diagrams for all density grids with respect to the reference phase diagram computed on the default initial spatial condition setup (a bi-centric symmetrical non toroidal configuration) using the measure defined in equation~\ref{eq:phase-distance}. For each density grid, we obtain the average squared distance between corresponding points of the phase diagrams, i.e. the mean value of the final output measure, such as segregation or inequality, for a given value of parameters in the two setups (isotropic and generated). This average squared distance for each point is then related to the average variance of each of the phase diagram (the reference one and the one for the grid under inquiry). Therefore, values greater than 1 will mean that inter-diagram variability is more important than intra-diagram variability. We tested the sensitivity to the type of distance, using a Minkovski distance with a varying exponent. The results are presented in~\ref{app:distances}, and show a similar sensitivity to spatial initial conditions.

\subsubsection{Sugarscape}

% summary stats
%   Min. 1st Qu.  Median    Mean 3rd Qu.    Max. 
% 0.08909 0.19790 1.52200 1.29600 2.16400 2.98100 

\begin{figure}[!t]
\centering
	\includegraphics[width=0.8\textwidth]{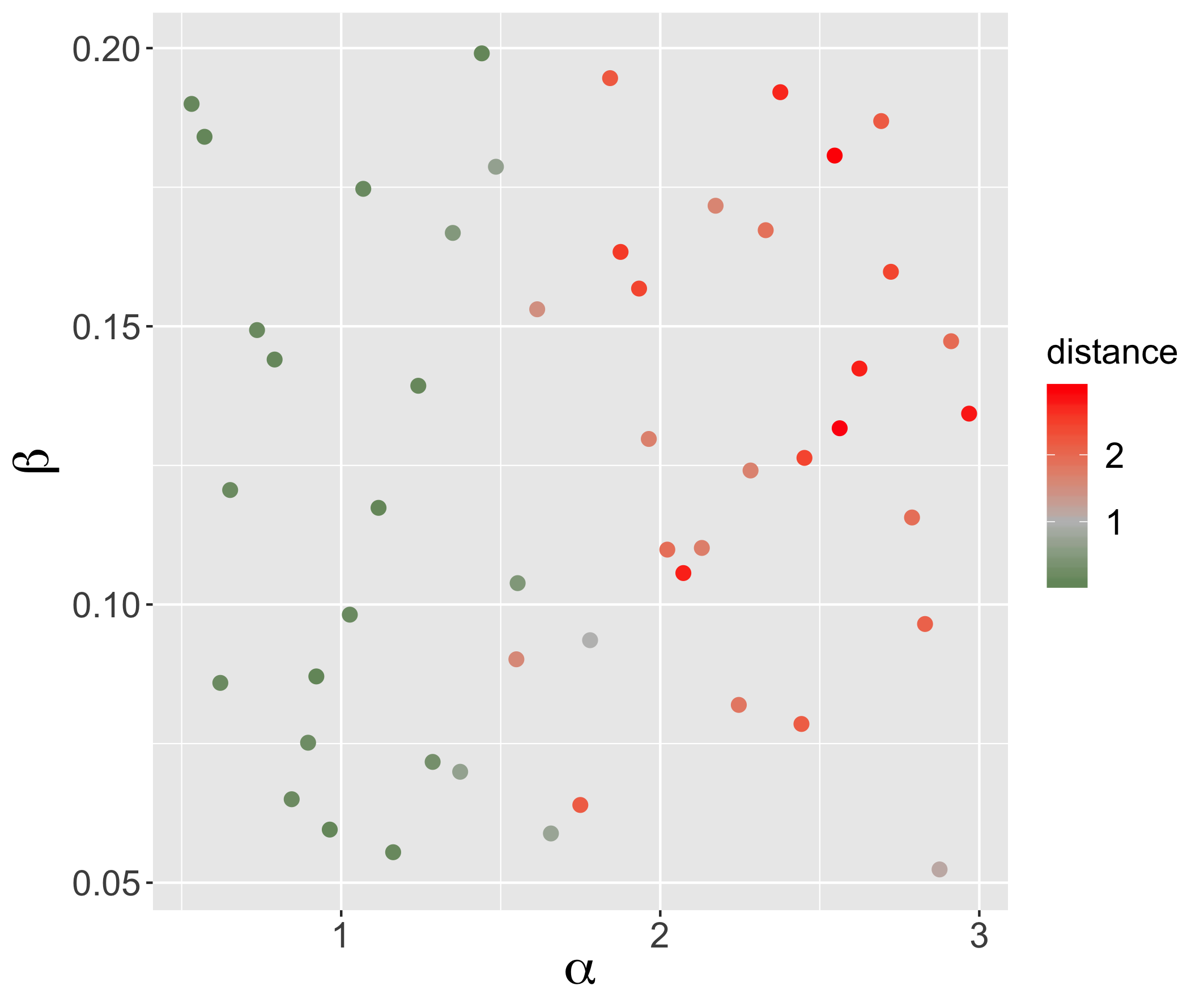}
\caption{\textbf{Relative distances of phase diagrams by initial spatial grids described by their generator parameters.} Relative distance as a function of generator parameters $\alpha$ (strength of preferential attachment) and $\beta$ (strength of diffusion process).}
\label{fig:sugarscape-distance-meta}
\end{figure}

We obtain a very strong sensitivity to initial spatial conditions for the Sugarscape model. Indeed, the relative distance between the phase diagrams of different density grids and the phase diagram of the reference case ranges from 0.09 to 2.98 with a median of 1.52 and an average value of 1.30. The mean distance above 1 means that, on average, the model is more sensitive to the generator parameters than to its own parameters (population and sugar endowment) in the reference model. Moreover, the maximum distance of 2.98 means that the variation due to the change of grid can be up to three times bigger than the variation due to the model parameters. We plot in Fig.~\ref{fig:sugarscape-distance-meta} the distribution of these distances in the generator parameter space. Each point represent one of the 50 different density grids used to initialise the distribution of sugar in the model. The points are projected with respect to the generator parameters, and coloured according to the relative distance of the phase diagram of the simulations using this grid to the phase diagram of the reference case. Therefore, the figure ~\ref{fig:sugarscape-distance-meta} shows that the grids generated with a high $\alpha$ (i.e. with a small number of very high density cells) produce simulation results that vary more between the reference case and the generated grid with the same values of parameters than within the reference case because of parameter variations. This pattern is emphasized when grids are generated with a high $\alpha$ and a high $\beta$ (i.e. with low gradient of density decrease around the kernels of high density). These grids have the highest relative distance to the reference case. On the contrary, with grids closer to the uniform pattern of the reference case (bottom left of the graph), the model parameters are more important in determining the final inequality levels than the initial spatial distribution of sugar.

%%%%%%%%%%%%%
%%%%%%%%%%%%%

\begin{figure}[!t]
\centering
	\includegraphics[width=0.8\textwidth]{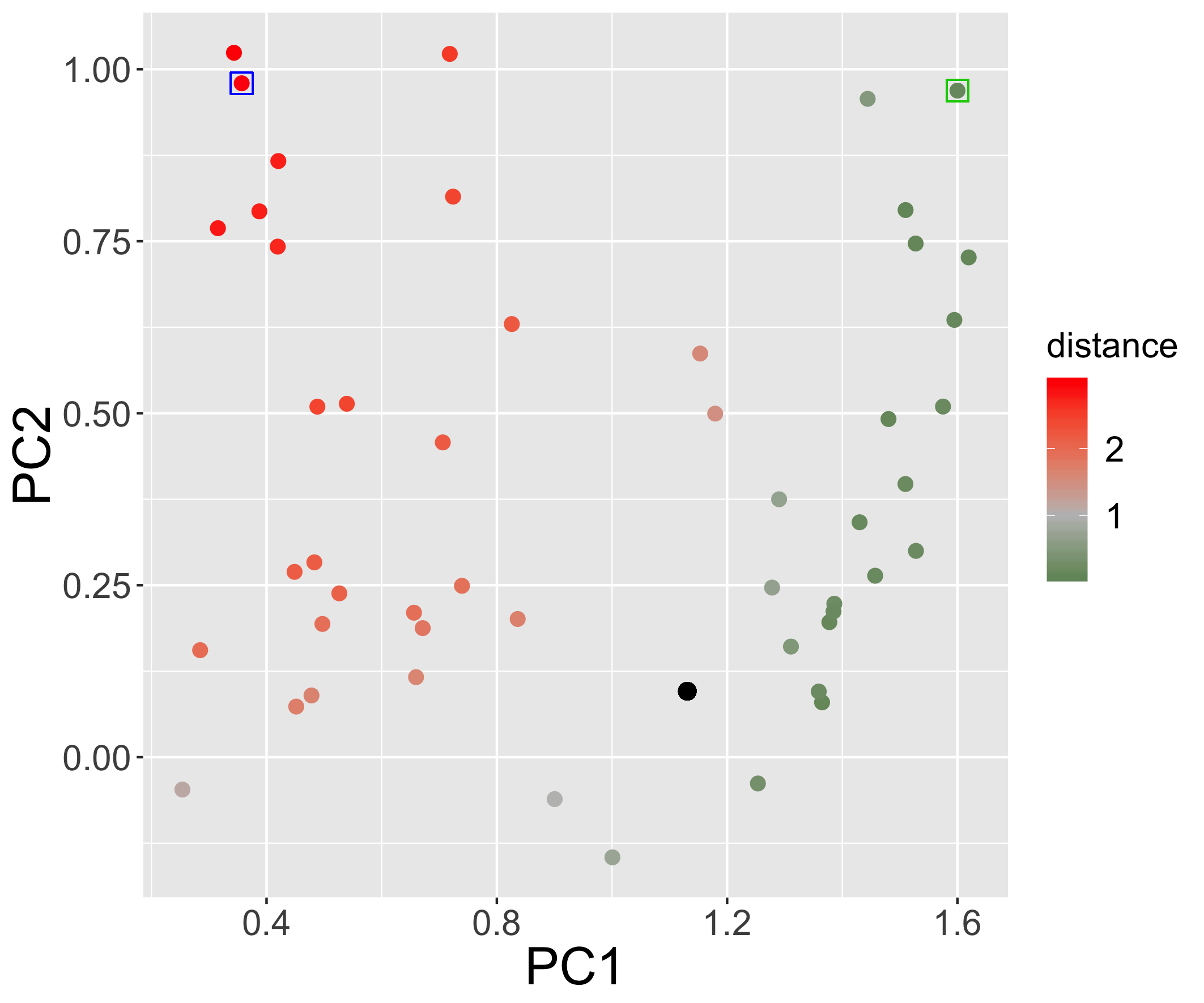}
\caption{\textbf{Relative distances of phase diagrams to the reference across grids.} Relative distance as a function of two first principal components of the morphological space (see text). Black point correspond to the reference spatial configuration. Green frame and blue frame give respectively the first and second particular phase diagrams shown in Fig.~\ref{fig:sugarscape-phasediagrams}.}
\label{fig:sugarscape-distance-pca}
\end{figure}

Another way of quantifying the density grids, instead of looking at the generator parameters, is to look at the resulting indicators of urban form, such as Moran's I, average distance, rank-size slope and entropy (see~\cite{LeNechet2015} for precise definition and context). This 4-dimensional space defined a morphological space. For the purpose of interpretability and visualisation, we reduce this space to a bi-dimensional space with a principal component analysis. The first two components represent 92\% of cumulated variance. The first component defines a ``level of sprawl'' and of scattering, whereas the second one represents the level aggregation.\footnote{We have $PC1 = 0.76\cdot distance + 0.60\cdot entropy + 0.03\cdot moran + 0.24\cdot slope$ and $PC2 = -0.26\cdot distance + 0.18\cdot entropy + 0.91\cdot moran + 0.26\cdot slope$.} We find that grids producing the highest deviations are the ones with a low level of sprawl and a high aggregation (top left of figure \ref{fig:sugarscape-distance-pca}). It is confirmed by the behaviour as a function of generator parameters, as high values of $\alpha$ also yield high distance. In terms of model processes, it shows that congestion mechanisms in the gathering of the resource induces fast increases of inequality. To put these results in perspective of our workflow given in Fig.~\ref{fig:method}, we have a sensitivity to spatial parameters on average greater than the sensitivity to model parameters.

% pca of morphological space
% "","PC1","PC2","PC3","PC4"
%"distance",0.762358566609464,-0.260991693298744,0.200656405132039,0.557162237616392
%"entropy",0.601306167355116,0.181706245959277,0.0958379422351422,-0.772158547261002
%"moran",0.0311129390452153,0.912155429075071,0.30114271129527,0.276256268103684
%"slope",0.237217819823539,0.258531718397015,-0.927289147645628,0.130475642169329

%%%%%%%%%%%%%
%%%%%%%%%%%%%

\subsubsection{Schelling} 
Within a standard Schelling model (i.e. initialised with a uniform density grid), \citet{Gauvinetal2009} have built the phase diagram of segregation patterns depending on the combination of parameter values. For high levels of tolerance ($S < 0.25$), there is no segregation. For high values of vacancies ($V > 0.65$) and low values of tolerance ($S > 0.5$), these is a diluted segregation state where homogeneous communities are separated from other by large empty buffers. Finally, for low values of vacancies ($V < 0.2$) and low values of tolerance ($S > 0.7$), the model is frozen in a state where everyone is unhappy but no-one can express their intolerant behaviour due to the lack of free spaces. Between these extreme cases, the model gives rise to segregated states where homogeneous communities adjoin one another. The objective of this quantitative experiment is to evaluate to which extent this phase diagram is modified when different density grids are applied. We show in Fig.\ref{fig:schelling-distance-meta} (Supplementary Material) the values of the relative distance as a function of meta-parameters and in the reduced morphological space, in a way similar to the analyse done with Sugarscape. Variations are less considerable than for Sugarscape across phase diagrams, but values close to 1 show that several configurations are as sensitive to initial spatial conditions than to their parameters. We focus in the following on a qualitative characterisation of these variations.

\subsection{Qualitative variations}
\label{sec:qualResults}

\subsubsection{Schelling}

In this qualitative exploration of the effect of initial spatial conditions on the results of Schelling's model, we use the classification of grids into three morphological types (cf. figure \ref{fig:densityTypes}). In particular, we want to evaluate to which extent the typology summarises the spatial effects, and if one type of urban form or another enhances the segregation mechanism of the model, or interacts differently with the model parameters. This experiment attempts at drawing conclusions on urban morphology, beyond the technical conclusions already obtained with respect to simulation sensitivity.

\begin{table}[]
\centering
\begin{threeparttable}
\caption{\textbf{Regression of the segregation level of Schelling simulation with order parameters and type of city grid.}}
\label{tab:regressionSchelling}
%\begin{adjustwidth}{-1cm}{-1cm}
\begin{tabular}{|p{2.5cm}|ll|ll|ll|}
\hline
Simulation outcome by segregation index:    & \multicolumn{2}{c|}{\textbf{Dissimilarity}}   & \multicolumn{2}{c|}{\textbf{Entropy}} & \multicolumn{2}{c|}{\textbf{Moran's I}} \\ \hline
\textbf{Intercept}                          & -0.212 *** & -0.141 ***                       & -0.254 ***        & -0.208 ***        & -0.036 ***           & -0.061 ***               \\ \hline
\textbf{Similarity Wanted (S)}              & 1.212 ***  & 1.212 ***                        & 1.250 ***         & 1.250 ***         & 0.550 ***            & 0.550 ***                \\ 
\textbf{quadratic term ($S^2$)}               & -0.942 *** & -0.942 ***                       & -0.963 ***        & -0.963 ***        & -0.428 ***           & -0.438 ***               \\ 
\textbf{Vacancy Rate (V)}                   & 0.602 ***  & 0.602 ***                        & 0.453 ***         & 0.453 ***         & -0.027 ***           & -0.027 ***               \\ 
\textbf{Minority Index (\%Maj - \%Min)}     & 0.307 ***  & 0.307 ***                        & 0.130 ***         & 0.130 ***         & -0.067 ***           & -0.067 ***               \\ \hline
\textbf{Density Grid = Polycentric}         &            & 0.087 ***                        &                   & 0.052 ***         &                      & 0.001 ***                \\ 
\textbf{Density Grid = Discontinuous}       &            & 0.111 ***                        &                   & 0.068 ***         &                      & \textit{0.00}              \\
\textbf{Attraction generator parameter $\alpha$} &            & -0.083 ***                       &                   & -0.053 ***        &                      & 0.014 ***                \\ 
\textbf{Diffusion generator parameter $\beta$}   &            & 0.323 ***                        &                   & 0.218 ***         &                      & 0.017 ***           \\ \hline
\textbf{R2 (\%)}                            & 30.6       & 34.7                             & 24.1              & 25.6              & 23.9                 & 24.0                    \\ 
\textbf{\# of observations (sim. runs)}     & 2,106,000  & 2,106,000 						 & 2,106,000          & 2,106,000          & 2,106,000             & 2,106,000                \\ 
\textbf{AIC}                                & -70717.68   & -198748.2  						& 208213.8          & 166048.8          & -4385990             & -4387816                 \\ \hline
\end{tabular}
%\end{adjustwidth}
\begin{tablenotes}
 \item Moran's I applies to the minority population
 \item *** means that the estimate is significant at the 0.01 level.
\end{tablenotes}
  \end{threeparttable}
\end{table}

In table \ref{tab:regressionSchelling}, we see that the type of density grid with which the model is initialised correlates to a certain extent with the level of segregation measured at the end of the simulation run. Indeed, compared to the reference case of compact (monocentric) density patterns, polycentric grids produce more dissimilarity and entropy between the location of green and red agents. Discontinuous grids have the same effect, although attenuated. The results obtained with Moran's I are opposite, because this index measures spatial autocorrelation at the global level and that compact cities have higher levels of global autocorrelation by construction. However, linear models with and without the type of density distribution yield the same coefficients for Schelling's parameters $V$ and $S$, the only exception being the vacancy rate $V$ in the Moran's I model with grid types, which becomes non-significant. The similarity of the coefficient in both cases means that the effect of the model's parameters (and thus the mechanism by which agents of similar group cluster in space) is the same regardless of the distribution of density. The way polycentric and discontinuous density grid exhibit higher segregation is by allowing buffer zones of low density to surround pockets of homogeneity, which is impossible in a compact city, because everyone is at reach of everyone else. The buffering process confirms previous results obtained with network structures \citep{Banos2012} and supports the conclusion that space acts here on top of mechanisms rather than in interaction with them.

\subsubsection{Sugarscape}

\begin{figure}[!t]
\centering
	\includegraphics[width=\textwidth]{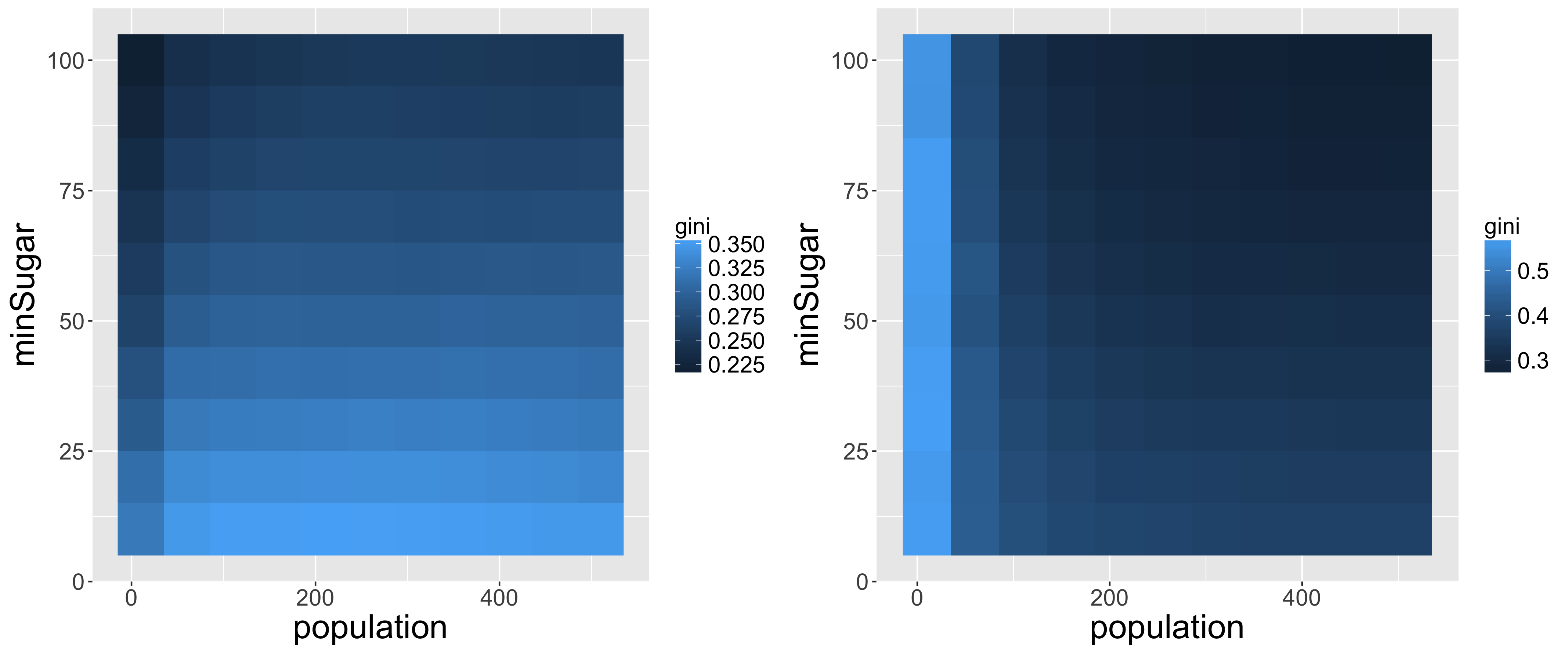}
\caption{\textbf{Examples of phase diagrams.} We show two dimensional phase diagrams on $(P,s_-)$, both at fixed $s_+ = 110$. (Left) Green frame, obtained with $\alpha = 0.79$, $n=2$, $\beta = 0.14$, $N=157$; (Right) Blue frame, obtained with $\alpha = 2.56$, $n=3$, $\beta = 0.13$, $N=128$.}
\label{fig:sugarscape-phasediagrams}
\end{figure}

We now check the sensitivity in terms of qualitative behavior of phase diagrams. We show the phase diagrams for two very opposite morphologies in terms of sprawling, but controlling for aggregation with the same $PC2$ value. These correspond to the green and blue frames in Fig.~\ref{fig:sugarscape-distance-pca}. In terms of grid shape, we observe that the difference between the two grids is mainly on average distance and entropy: in a nutshell, the first grid is much more dispersed and disorganised than the second. Although the behaviours are rather stable for varying $s_+$, the initial maximum endowment in sugar, which means that the poorest agents have a determinant role in trajectories, the two examples have not only a very distant baseline inequality (the ceiling of the first 0.35 is roughly the floor of the second 0.3), but their qualitative behavior is also radically opposite: the sprawled configuration gives inequalities decreasing as population decreases and decreasing as minimal wealth increases, whereas the concentrated one gives inequalities strongly increasing as population decreases and also decreasing with minimal weights but significantly only for large population values (fig. \ref{fig:sugarscape-phasediagrams}. In sprawled spaces, inequalities are thus fostered by a lack of minimal local resources, whereas population will drive inequality in concentrated spaces. The process is thus completely reversed depending on the grid chosen to run the model on, which would have significant impacts if one tried to deduce policy from this model.
%This second example confirms thus the importance of sensitivity of simulation models to the initial spatial conditions.

% phase diagrams -> ok well different qualitatively
%          spAlpha spDiffsteps spDiffusion spGrowth spPopulation
% id=27 : 0.7913103    2.376837   0.1440293 157.4147 4852.746
% id=0 : 2.562398    3.753032   0.1316788 128.4632 4753.983
% maxSugar = 110

%%%%%%%%%%%%%
%%%%%%%%%%%%%

%%%%%%%%%%%%%%%%%%%%%%
\section{Discussion}
%%%%%%%%%%%%%%%%%%%%%%

We consider that the method presented in this paper holds great potential for strengthening geographical models' exploration. However, two limits and two areas of opportunities have still not been tackled. 

\subsection{Limits}

\subsubsection{Comparing phase diagrams}

Comparing phase diagrams is as we saw not so straightforward, and further developments of our method imply testing alternative methods for this particular point. For example in the case of the Schelling model, an anisotropic spatial segregation index (giving the number of clusters found and in which region in the parameter spaces they are roughly situated) would differentiate strong phase transitions in the space of generator parameters. The use of metrics comparing spatial distributions, such as the Earth Movers Distance which is used for example in Computer Vision to compare probability distributions~\citep{rubner2000earth}, or the comparison of aggregated transition matrices of the dynamic associated to the potential described by each distribution, would also be potential tools. Map comparison methods, popular in environmental sciences, provide numerous tools to compare two dimensional fields~\citep{visser2006map,kuhnert2005comparing}. To compare a spatial field evolving in time, elaborated methods such as Empirical Orthogonal Functions that isolates temporal from spatial variations, would be applicable in our case by taking time as a parameter dimension, but these have been shown to perform similarly to direct visual inspection when averaged over a crowdsourcing~\citep{10.1371/journal.pone.0178165}. The transfer of methods used to compare sequences \citep{kruskal1983overview} or time-series \citep{liao2005clustering} is a possible way to develop measures between phase diagrams. The higher dimension of the phase diagrams we study must however be considered with caution when transferring methods, in a way analog to the application of global sensitivity indexes to spatial data \citep{lilburne2009sensitivity}. We can also note than more generally, this problem of comparing phase diagrams is a particular instance of the more generic issue of comparing patterns, which for example include unsupervised learning techniques \citep{hastie2009unsupervised}. The investigation of diverse approaches to systematically quantify differences between phase diagrams is an important potential development of our method.

\subsubsection{Platform constraints and docking challenges}

An aspect that we have not touched upon in the article with respect to the sensitivity to initial spatial conditions is the importance of the modelling platform as a constraint in the formalisation of space. For example, spatial structure may be easier to implement as a raster rather a vector in NetLogo models, which could influence the implementation choices of some non-experienced modelers. Its toroidal default setting might also have influenced the work of many modellers who did not question explicitly the representation of space. This issue is part of the docking challenge \citep{Axtelletal1996} (i.e. checking if two models can produce the same results), but more generally, it involves a description of the model and its spatial requirements more detailed than what is currently the rule.

\subsection{Opportunities and extensions}

\subsubsection{Reproducibility and Applicability}

Although the applications we present here are limited by the simplicity of the models, we think that the method could (and should) be applied to larger models including domain mechanisms and more empirical initialisation data, for example synthetic populations. The sensitivity analysis to initial spatial conditions could then be either a replication on the spatial allocation of the synthetic population, or a series of spatial permutations of the empirical spatial inputs.
We want to foster this extension of our work by releasing the density grids also generated, as well as the generating work-flow and the model implementation. They are available on the open repository of the project at \url{https://github.com/AnonymousAuthor2/SpaceMatters}. Future work could be done to compare these or generate grids with a larger morphological span, covering other typical urban forms that can be found in the world.

Another way to go would be to implement additional generators, such as social networks \citep{alizadeh2016generating} with localised agents, transportation networks generators \citep{raimbault2018multi}, or coupled road network and population raster generators \citep{raimbault2018urban}.
% Generators of abstract urban systems at a macroscopic scale could also be useful tools to better understand urban growth models as sketched by \cite{raimbault2018unveiling}.}  

\subsubsection{An emancipation opportunity for social sciences.}

As~\citet{pumain2003approche} points out in an overview of complexity approaches in geography, transfer of models and concepts between disciplines may induce a transfer of corresponding assumptions. Geography and the social sciences in general have been strongly influenced by physics in the last decades, that beside their highly enriching impact~\citep{o2015physicists}, may have softly imposed strong assumptions such as homogeneity and isotropy of space in basic models. We believe that a renewed approach on the role of space as we proposed, in other terms insisting that \emph{space matters}, is an opportunity the for social sciences to build their own stream of methodologies in the modelling domain.

\section{Conclusion}

After reviewing the extensive literature on spatial biases in statistical and simulation models, we presented a method to analyse the sensitivity of a simulation's results to the initial spatial configuration. We did so by implementing a spatial generator whose output is used as input for the simulation model. We applied this approach to two textbook ABMs: Schelling and Sugarscape. With the Schelling experiment, we found that the different urban morphologies impact the interaction patterns, and that polycentric and discontinuous cities appear systematically more segregated than compact cities in terms of dissimilarity and entropy index. With Sugarscape, we show that the model is more sensitive to space than to its other parameters in the reference Netlogo implementation, both qualitatively and quantitatively: the amplitude of variations across density grids is larger than the amplitude in each phase diagram, and the behaviour of the phase diagram is qualitatively different in different regions of the morphological space. We think that this method has the potential to increase the arsenal of evaluation of geographical models, in order to assess the sensitivity of models to their initial spatial conditions but also to learn about the impact of the urban form on social mechanisms.
%%%%%%%%%%%%%
% Acknowledgements

\section*{Acknowledgements}

The authors acknowledge the funding of their institutions and the EPSRC project number EP/M023583/1. Results obtained in this paper were computed on the vo.complex-system.eu virtual organization of the European Grid Infrastructure ( http://www.egi.eu ). We thank the European Grid Infrastructure and its supporting National Grid Initiatives (France-Grilles in particular) for providing the technical support and infrastructure.

\section*{Appendix A: Behavior of the density grid generator}

\begin{figure}[!t]
	\centering
    \includegraphics[width=\textwidth]{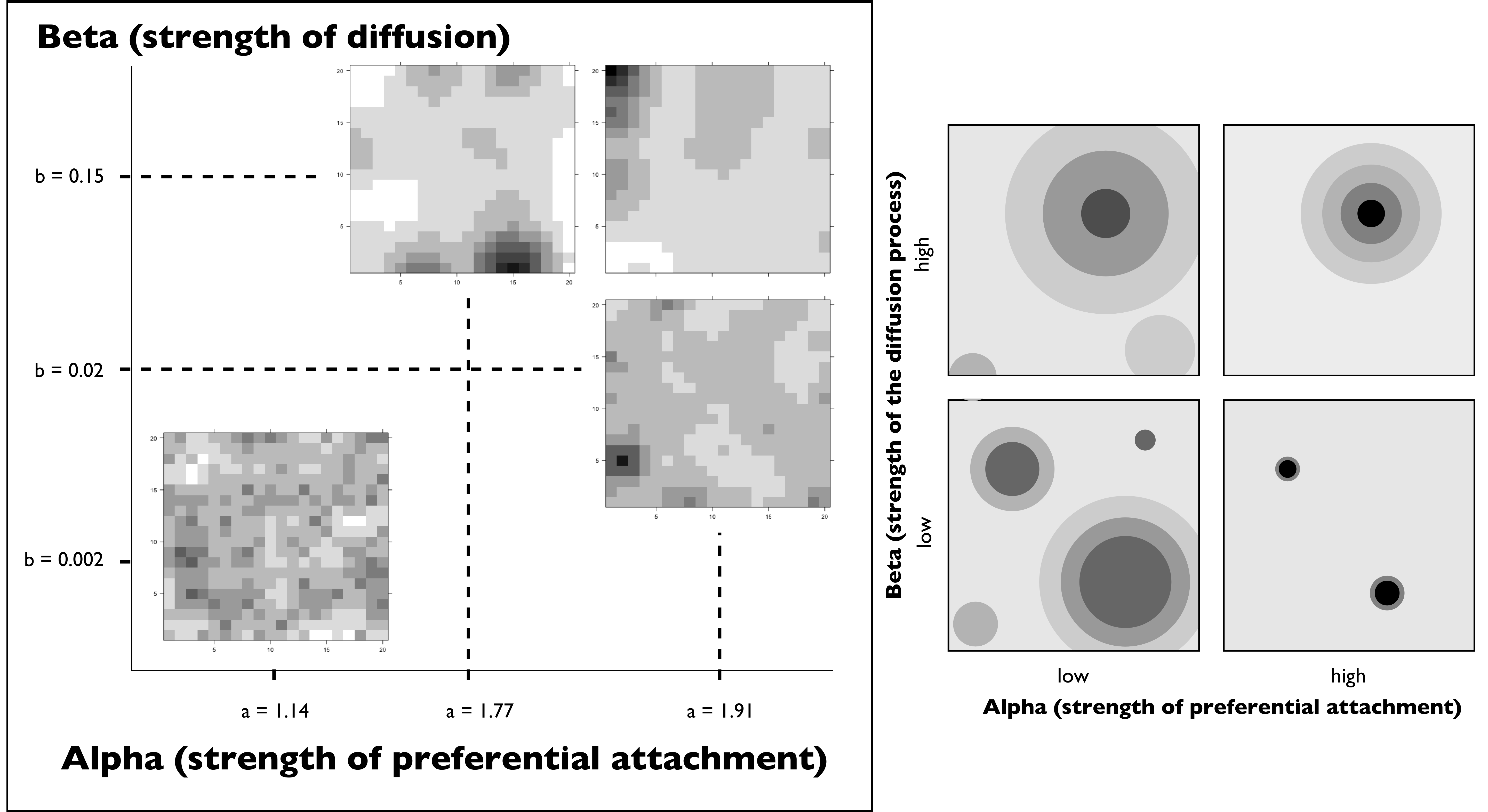}
    \caption{\textbf{Stylized behavior of the density grid generator.} (Left) Examples of grids with their position in the $(\alpha,\beta)$ parameter space; (Right) Stylized interpretation of the forms obtained as a function of $\alpha$ and $\beta$.}\label{fig:stylized-generator}
\end{figure}

We give in Fig.~\ref{fig:stylized-generator} plots roughly summarizing the behavior of the density grid generator, according to parameters $(\alpha,\beta)$. To put it in a simple way, high values of $\alpha$ give highly hierarchical configurations, and diminishing $\beta$ increase the number of centers. Low values of $\alpha$ give diffuse patterns, with however clear centers for a high diffusion. We do not discuss here the role of other parameters, but according to \cite{raimbault2018calibration}, diffusion steps give smoother forms, and the rate between total population and the population increment at each step (which is equivalent to the total number of steps) is crucial to select non-stationary distributions that are closer to real configurations. \cite{raimbault2018calibration} also shows the existence of non-linear behaviors in some regions of the parameter space, so the description we gave here shall not be interpreted as a linear link between generator parameters and the morphological properties of the generated grids.

%%%%%%%%%%%%%%
%%%%%%%%%%%%%%

%%%%%%%%%%%%%%%%%
\section*{Appendix B: Additional statistical analysis}

\label{app:convergence}

%%%
%% Sugarscape

% Sharpes
%summary(sres$giniSharpe)
%   Min. 1st Qu.  Median    Mean 3rd Qu.    Max. 
%  2.343  14.413  20.993  20.098  26.208  49.808 

% # ks
% # table(sres$distrib)
% #gamma  geom lnorm  norm  unif 
% #1090  9794 16212 14263 13201 
% AIC
% table(sres$distrib)
% exp gamma lnorm  norm 
%  1451  9707 22825 20577 

%%%
%% Schelling
% Min. 1st Qu.  Median    Mean 3rd Qu.    Max. 
%  1.053  12.368  19.434  25.268  24.328 468.563 
%> summary(sres$entropySharpe)
%    Min.  1st Qu.   Median     Mean  3rd Qu.     Max. 
%  0.5458   8.2591  14.5112  17.0100  17.9372 244.3230 
%> summary(sres$moranSharpe)
%    Min.  1st Qu.   Median     Mean  3rd Qu.     Max. 
%  0.0001   0.5554   0.7782   3.2724   3.8979 121.6829 

For both models we estimate the Sharpe ratios for each indicator by $S(X) = \hat{\mathbb{E}}\left[X\right]/\hat{\sigma}(X)$ with standard estimators for average and standard deviation. The summary statistics of this ratio computed on repetitions for all parameter points are given in Table~\ref{tab:summarysharpes}. Under the assumption of a normal distribution, the width of the confidence interval at level $c$ is given by $\left|\mu_+ - \mu_-\right| = 4\cdot \sigma \cdot z_{c} / \sqrt{n}$ where $\sigma$ is the standard deviation, $z_{c}$ is the quantile at which $c$ is attained by the cumulative distribution, which is around $1.96$ for a 95\% confidence interval. This means that to obtain a confidence interval of width $\kappa \cdot \sigma$, one needs a number $n \simeq (8 / \kappa )^2$ of repetitions. This gives 64 repetitions for $\kappa = 1$. As the Sharpe ratios are in general smaller for Schelling indicators than for Sugarscape, we take $n = 50$ for Sugarscape and $n = 100$ for Schelling to have a similar confidence in estimations.

%%%%%%%%%%%%%%
\begin{table}[!t]
\centering
	\begin{tabular}{|c|cccccc|}
	\hline
Model/Indicator & Min. & 1st Qu. &  Median &  Mean & 3rd Qu. & Max. \\\hline
Sugarscape/Gini & 2.343 & 14.413 & 20.993 & 20.098 & 26.208 & 49.808\\\hline 
Schelling/Dissimilarity & 1.053 & 12.368 & 19.434 & 25.268 & 24.328 & 468.563\\
Schelling/Entropy & 0.5458 & 8.2591 & 14.5112 & 17.0100 & 17.9372 & 244.3230\\
Schelling/Moran & 0.0001 & 0.5554 & 0.7782 & 3.2724 & 3.8979 & 121.6829\\\hline
    \end{tabular}
    \caption{\textbf{Summary statistics of Sharpe ratio estimated on repetitions for each parameter point.}}
\label{tab:summarysharpes}
\end{table}
%%%%%%%%%%%%%%

%%%%%%%%%%%%%%%%%
\section*{Appendix C: Additional figures for the Schelling model}

\begin{figure}[!t]
\centering
\includegraphics[width=\textwidth]{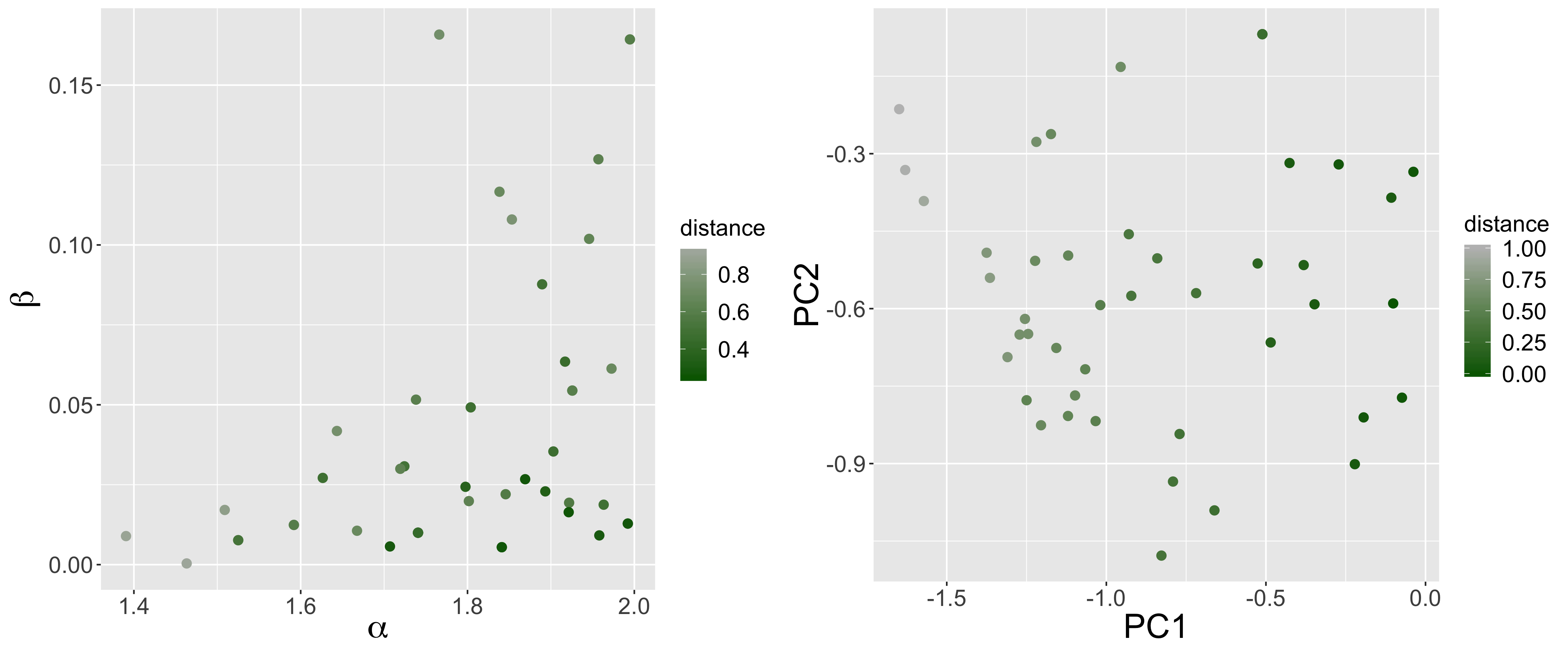}
\caption{\textbf{Relative distances of phase diagrams to the reference across grids for the Schelling model.} Each point corresponds to a spatial configuration and colour gives the relative distance to one of the phase diagrams. We show them in the generator parameter space (Left) and in the reduced morphological space (Right).\label{fig:schelling-distance-meta}}
\end{figure}

The Fig.~\ref{fig:schelling-distance-meta} gives the phase diagrams distances as a function of generator parameters and morphological components, similarly to the Sugarscape model in main text. In absolute, this version of the Schelling model seems less sensitive to density grids than the sugarscape model, as we do not obtain a high range of values here. We however obtain measures ranging from 0 to 0.85 with the euclidian distance, what is however characteristic of a significant sensitivity to space.

%summary(dists)
%   Min. 1st Qu.  Median    Mean 3rd Qu.    Max. 
% 0.0000  0.2164  0.3325  0.3700  0.4846  0.8473 

%%%%%%%%%%%%%
%%%%%%%%%%%%%

%%%%%%%%%%%%%%%
\section*{Appendix D: Comparison of phase diagram with other distances}

\label{app:distances}

We describe here the tests done with other distances to compare phase diagrams. We tested normalized Minkovski distances, defined by $d(x,y) = \left(\frac{1}{N}\cdot \sum_i \left|x_i - y_i\right|^{q}\right)^{\frac{1}{q}}$, for varying values of $q$ from $q = 1$ (Manhattan distance) to $q = 10$, including $q = 2$ (Euclidian distance) which is used in main text. The Table~\ref{tab:schellingdistances} gives the summary statistics of each distance computed on all initial configurations for the Schelling model. The Table~\ref{tab:sugarscapedistances} gives the same statistics for the sugarscape model. We naturally obtain smaller difference with the Manhattan distance but which remain significant (averages of 10\% for the Schelling model and 40 \% for sugarscape), and variabilities with higher values of the Minkovski exponent are much higher. This results confirm the high variability observed in main text with the Euclidian distance.

% Schelling
% [1] "alpha = 1"
%   Min. 1st Qu.  Median    Mean 3rd Qu.    Max. 
%0.00000 0.02838 0.06263 0.10608 0.16818 0.37018 
%[1] "alpha = 2"
%   Min. 1st Qu.  Median    Mean 3rd Qu.    Max. 
% 0.0000  0.1520  0.2517  0.3107  0.4359  0.8155 
%[1] "alpha = 3"
%   Min. 1st Qu.  Median    Mean 3rd Qu.    Max. 
% 0.0000  0.3709  0.5133  0.5860  0.7435  1.2930 
%[1] "alpha = 10"
%   Min. 1st Qu.  Median    Mean 3rd Qu.    Max. 
%  0.000   2.083   2.431   2.380   2.713   3.664 

% Sugarscape
%[1] "alpha = 1"
%   Min. 1st Qu.  Median    Mean 3rd Qu.    Max. 
%0.07184 0.13996 0.39816 0.42452 0.64751 1.13169 
%[1] "alpha = 2"
%   Min. 1st Qu.  Median    Mean 3rd Qu.    Max. 
%0.08909 0.19794 1.52272 1.29594 2.16371 2.98273 
%[1] "alpha = 3"
%   Min. 1st Qu.  Median    Mean 3rd Qu.    Max. 
% 0.1025  0.2288  3.1495  2.5074  4.2718  5.5378 
%[1] "alpha = 10"
%   Min. 1st Qu.  Median    Mean 3rd Qu.    Max. 
% 0.1540  0.4466  9.8757  7.6246 13.0550 16.8001 

\begin{table}[!t]
	
	\centering
	\begin{tabular}{|c|cccccc|}
	\hline
    	$q$ & Min. & 1st Qu. & Median & Mean & 3rd Qu. & Max. \\\hline
        1 & 0.00000 & 0.02838 & 0.06263 & 0.10608 & 0.16818 & 0.37018 \\
        2 & 0.0000 & 0.1520 & 0.2517 & 0.3107 & 0.4359 & 0.8155 \\
        3 & 0.0000 & 0.3709 & 0.5133 & 0.5860 & 0.7435 & 1.2930 \\
        10 & 0.000 & 2.083 & 2.431 & 2.380 & 2.713 & 3.664 \\\hline
    \end{tabular}
    \caption{\textbf{Summary statistics of different distances for schelling.}}\label{tab:schellingdistances}
\end{table}

\begin{table}[!t]
	
	\centering
	\begin{tabular}{|c|cccccc|}
	\hline
    	$q$ & Min. & 1st Qu. & Median & Mean & 3rd Qu. & Max. \\\hline
        1 & 0.07184 & 0.13996 & 0.39816 & 0.42452 & 0.64751 & 1.13169\\
        2 & 0.08909 & 0.19794 & 1.52272 & 1.29594 & 2.16371 & 2.98273\\
        3 & 0.1025 & 0.2288 & 3.1495 & 2.5074 & 4.2718 & 5.5378\\
        10 & 0.1540 & 0.4466 & 9.8757 & 7.6246 & 13.0550 & 16.8001\\\hline
    \end{tabular}
    \caption{\textbf{Summary statistics of different distances for sugarscape.}}\label{tab:sugarscapedistances}
\end{table}

%%%%%%%%%%%%%%%
\section*{Appendix E: Pseudo-code for models}
\label{app:code}

We give below the pseudo-code for the implementations we used of both sugarscape (Fig.~\ref{frame:sugarscapecode}) and Schelling (Fig.~\ref{frame:schellingcode}) models. We recall that source code is openly available at \url{https://github.com/AnonymousAuthor2/SpaceMatters}. The pseudo-code is in the style of NetLogo code, which is already easily readable.

%%%%%%%%%%%%%%%
\begin{figure}[!t]
\caption{\textbf{Pseudo-code (NetLogo style) for the Schelling model used here.} We give a pseudo-code very close to the NetLogo language, at the exception of italic expressions which are not put here in NetLogo to ease understandability.}\label{frame:schellingcode}
\bigskip
\begin{mdframed}
\begin{lstlisting}
while not stopping-condition [
	ask turtles [
		set unsatisfied? false
		let neighboragents turtles in-radius neighborhood-radius
		let wantedcolor color
		let stranger-rate count neighboragents
		   with [color = wantedcolor] / count neighboragents
		if stranger-rate < similar-wanted [
			set unsatisfied? true random-move
		]
	]
    \end{lstlisting}
    \end{mdframed}
\end{figure}
%%%%%%%%%%%%%%%

%%%%%%%%%%%%%%%
\begin{figure}[!t]
% Minimal netlogo code for the sugarscape model
\caption{\textbf{Pseudo-code (NetLogo style) for the sugarscape model.} Conventions are the same than for Schelling pseudo-code.}\label{frame:sugarscapecode}
\bigskip
\begin{mdframed}
\begin{lstlisting}
while [time < final-time] [
    ;regrow the patch resources
    ask patches [set psugar min (list max-psugar (psugar + 1))]
	
    ; make agents exploit resources
    ask turtles [
		; candidate patches to which the turtle can move
		let move-candidates patches in vision range with no turtles
		
		; select the patch with maximal resource
		let possible-winners move-candidates with-max [psugar]
		if any? possible-winners [
			move-to one-of possible-winners at minimal distance
		]
		
		; eat the sugar on the patch
		set sugar (sugar - metabolism + psugar)
		set psugar 0
		
		; age and die if necessary
		set age (age + 1)
		if sugar <= 0 or age > max-age [
			; hatch an other turtle before dying
			hatch 1 [ new-turtle ]
			die
		]
	]
]
\end{lstlisting}
\end{mdframed}
\end{figure}
%%%%%%%%%%%%%%%

%\bibliographystyle{apalike}
%\bibliography{spacematters.bib}

%%%%%%%%%%%%%%%%%%%%%%%%%%%%%%%%%%%%%%%%%%%%%%

\end{document}